\documentclass[usenatbib]{mn2e}

\usepackage{graphicx}
\usepackage[hypertex]{hyperref}

\def \eg {e.g.}
\def \ie {i.e.}
\def\spose#1{\hbox to 0pt{#1\hss}}
\def\ltsim{$\mathrel{\spose{\lower 3pt\hbox{$\sim$}}
        \raise 2.0pt\hbox{$<$}}$\thinspace}
\def\gtsim{$\mathrel{\spose{\lower 3pt\hbox{$\sim$}}
        \raise 2.0pt\hbox{$>$}}$\thinspace}
\newcommand{\thin }{\thinspace}

\newcommand{\reff}{${\rm R_e}$}

\newcommand{\chandra }{{\em Chandra}}

\newcommand{\ned}{{\em{NED}}}

\newcommand{\mvir}{$M_{vir}$}
\newcommand{\cvir}{$c_{vir}$}
\newcommand{\hnought}{$H_0$}
\newcommand{\omegam}{$\Omega_m$}
\newcommand{\omegalambda}{$\Lambda$}
\newcommand{\xmm }{{\em XMM}}

\newcommand{\rosat }{{\em Rosat}}

\newcommand{\sigmac}{$\sigma_0$}

\defcitealias{humphrey06a}{H06}
\defcitealias{humphrey08a}{H08}
\defcitealias{humphrey09d}{H09}

\title[The slope of the mass profile in early-type galaxies]{The slope of the mass profile and the tilt of the fundamental plane in early-type galaxies}
\author[P.~J. Humphrey and D.~A. Buote]{Philip J. Humphrey and David A. Buote \\ Department of Physics and Astronomy, University of California, Irvine, 4129 Frederick Reines Hall, Irvine, CA 92697-4575}
\begin{document}
\date{Accepted 2009 December 26. Received 2009 December 9; in original form 2009 October 28}
\maketitle
\begin{abstract}
We present a survey, using the \chandra\ X-ray observatory, of the central gravitating mass profiles in a sample of 10 galaxies, groups
and clusters, spanning $\sim$2 orders of magnitude in virial mass.
We find the total mass distributions from $\sim$0.2--10\reff, where \reff\ is the optical
effective radius of the central galaxy, are remarkably similar to powerlaw density profiles.
The negative logarithmic slope of the mass density profiles, $\alpha$, systematically varies with \reff,
from $\alpha \simeq$2, for systems with \reff$\sim$4~kpc to $\alpha \simeq$1.2 for
systems with \reff\gtsim 30~kpc. Departures from hydrostatic equilibrium
are likely to be small and cannot easily explain this trend.
We show that the conspiracy between the baryonic (Sersic) and dark matter (NFW/ Einasto) 
components required 
to maintain a powerlaw {\em total} mass distribution naturally predicts an anti-correlation
between $\alpha$ and \reff\ that is very close to what is observed.
{The systematic variation of $\alpha$ with \reff\ }
implies a dark matter fraction within \reff\ that varies systematically
with the properties of the galaxy in such a manner as to reproduce, without fine
tuning, the observed tilt of the fundamental plane. 
We speculate that establishing a nearly powerlaw total mass distribution is therefore a fundamental 
feature of galaxy formation and the primary factor which determines the tilt
of the fundamental plane.
\end{abstract}

\section{Introduction} \label{sect_introduction}
The global optical properties of giant elliptical galaxies are often parametrized by three
key quantities, the effective (half-light) radius (\reff) of the stellar light, the mean
surface brightness (or luminosity) and the central line-of-sight velocity dispersion (\sigmac).
In this three-dimensional parameter space, they occupy a narrow ``fundamental'' plane
\citep{dressler87a,djorgovski87a}, the orientation of which differs significantly from 
naive expectations from the virial theorem
(assuming homology and a constant mass-to-light ratio, $\gamma$). 
The relative importance which deviations from homology, variations in the stellar population
properties and changing dark matter fractions play in producing this ``tilt'' have been hotly
debated
\citep[\eg][and references therein]{renzini93b,hjorth95a,ciotti96a,prugniel97a,graham97a,gerhard01a,padmanabhan04a,trujillo04a,cappellari06a,bolton07a,tortora09a}, with recent work tending to emphasize the importance
of dark matter \citep[\eg][]{cappellari06a,bolton07a}. To maintain the thinness of the 
fundamental plane, however, the dark matter fraction within $\sim$\reff\ must be tightly
correlated with the optical properties of the galaxy \citep[\eg][]{ciotti96a}.

Strong observational evidence for the existence of massive dark matter halos around
early-type galaxies has been provided by independent observational constraints from 
X-ray studies, lensing and stellar dynamics (\eg\ \citealt{humphrey06a}, hereafter \citetalias{humphrey06a};
\citealt{mathews03a} and references therein; \citealt{kochanek95a,griffiths96a,gavazzi07a,gerhard01a,thomas07a}).
Although the total mass distribution within \reff\ is dominated by the stars, the dark
matter fraction is non-negligible, and stellar dynamics and lensing studies (most of which are
restricted to within $\sim$\reff) suggest that
it establishes a conspiracy with the luminous matter to produce total mass density 
($\rho_m$) profiles close to  $\rho_m \propto R^{-2}$ (where R is the radius), 
\ie\ the ``singular isothermal sphere'' \citep[\eg][]{kochanek95a,treu04a,koopmans06a,koopmans09a,kronawitter00a}.
This ``bulge-halo conspiracy'' is similar to that which establishes 
flat rotation curves in disk galaxies.

Total mass distributions approximately consistent with $\rho_m \propto R^{-2}$ have
long been reported from hydrostatic X-ray analysis 
of nearby elliptical galaxies at scales much larger than \reff\
\citep[\eg][]{trinchieri86a,thomas86a,serlemitsos93a,nulsen95a,kim95a,rangarajan95a,matsushita98a}
and, in their \rosat\ imaging analysis of two galaxies, \citet{buote94,buote98d}
actually found an isothermal sphere potential to be preferred over some
other mass distributions. With the improved capabilities of \chandra\ and \xmm,
much tighter constraints on radial mass profiles have now been reported for a 
wider array of galaxies 
\citep[\eg][]{humphrey06a,humphrey08a,humphrey09d,fukazawa06a,osullivan04b,osullivan07b,buote02a},
which similarly resemble $\rho_m \propto R^{-2}$ \citep[as pointed out by][]{gavazzi07a},
although not exactly so over all radial scales \citep{romanowsky09a}.
\citet{fukazawa06a} fitted a model of the form $\rho_m \propto R^{-\alpha}$ 
to the combined mass profiles of their galaxy sample outside 10~kpc, finding 
$\alpha=1.67\pm0.33$.

At the scale of massive galaxy clusters, hydrostatic X-ray analysis has revealed 
much less cuspy mass distributions
\citep[$\alpha \sim$1--1.4; \eg][]{buote04b,voigt06a,arabadjis02a,david01a}.
In particular, \citet{lewis03a} and \citet[][]{zappacosta06a}
studied the very relaxed clusters A\thin 2029 and A\thin 2589, finding that the
{\em total} mass profiles were in good agreement with the NFW shape
expected for the dark matter {\em only}, once again suggesting 
some kind of conspiracy between the luminous and dark matter to produce an 
approximately powerlaw {\em total} mass distribution in the core of the system.
Stellar dynamics and lensing studies of clusters have, similarly, found less
cuspy total mass profiles than would be expected for an unmodified NFW component
plus the stellar mass \citep[\eg][]{sand04a,kelson02a}.

In order to investigate this disparate behaviour at different mass scales,
in this paper we carry out a uniform X-ray analysis of a sample of
relaxed galaxies,
groups and clusters to investigate the shape of the innermost mass distributions.
X-ray analysis is ideally suited for this study, since it allows
straightforward mass measurements in individual systems over a wide radial
range, from the baryon dominated regime (R\ltsim \reff) to 
regions where the dark matter dominates the gravitating mass (R$\gg$\reff),
thus providing adequate leverage to elucidate any conspiracy between these 
different components. For relaxed systems, hydrostatic equilibrium is 
believed to be an excellent approximation, with nonthermal effects contributing no
more than $\sim$20\%\ of the total pressure 
\citep[\eg][hereafter \citetalias{humphrey09d}]{churazov08a,nagai07a,piffaretti08a,humphrey09d}.
In our previous papers (\citetalias{humphrey06a,humphrey09d}; 
\citealt{gastaldello07a,zappacosta06a}) we have 
carried out detailed decompositions of the mass distribution of
these systems into baryonic and non-baryonic components. 
We have not, however, investigated in detail the relationship between these two components.
In light of the apparent conspiracies between them at both cluster and galaxy scales, 
in this present paper, we adopt the more pragmatic approach of examining whether
the {\em inner} part of the mass profile can be modelled as a powerlaw, and 
investigating whether its slope varies systematically.

Throughout this work, we assume a cosmology of \hnought=${\rm 70\ km\ s^{-1}\ Mpc^{-1}}$,
\omegam=$0.3$ and \omegalambda=0.7. All error-bars, unless stated otherwise, correspond to 
1-$\sigma$ confidence regions.


\section{Data Analysis}
\begin{table}
\caption{Properties of the galaxies, groups and clusters in our sample. We list the 
K-band effective radius of the central galaxy (\reff) obtained from the 2MASS database
\citep{jarrett00a}, except for the clusters, for which we report ($^\dag$) the V-band
\reff\ from \citet{malumuth85a} and ($^\ddag$) the R-band \reff\ from \citet{uson91a}. 
We quote luminosity distances for each object, based on the redshift in the \ned\ database,
except those marked ``$^*$'', which were based on the surface brightness fluctuations study
of \citet[][see \citetalias{humphrey06a,humphrey09d}]{tonry01}.
Deprojected density and temperature profiles were taken from the listed reference (Ref). 
Where the data are unpublished or 
were re-reduced in the present work, we list the Chandra observation identifiers (ObsID) 
and net exposure time (Exp).
References: (1) \citetalias{humphrey09d}, (2) \citet{humphrey09c}, (3) \citet{humphrey08a}, 
(4) F.\ Gastaldello (2009, priv.\ comm.); \citet{gastaldello07a}, (5) W.\ Liu (2009, priv.\ comm.); \citet{liu09a}} \label{table_sample}
\begin{tabular}{lllllr}
Object & \reff & Distance & ObsID & Exp & Ref \\
 & (kpc) & (Mpc) & & (ks) \\ \hline
NGC\thin 1332 & 2.7 & 21.3$^*$ & \ldots & \ldots & (1) \\
NGC\thin 720  & 3.1 & 25.7$^*$ &  7062,7372 &  99 & (2) \\
              &     &      &  8448,8449 &     &    \\
NGC\thin 4649 & 3.2 & 15.6$^*$ & \ldots & \ldots & (3) \\
NGC\thin 4261 & 3.4 & 29.3$^*$ & \ldots & \ldots & (1) \\
RXJ1159+5531  & 9.8 & 368  & \ldots & \ldots & (4) \\
MKW4          & 10 & 87   &  3234 & 30 & \ldots \\
AWM4          & 10 & 139  &   9423 & 71 & \ldots \\
ESO552-020    & 16 & 138  &   3206 & 19 & \ldots \\
ABELL\thin 2589 & 33$^\dag$ & 183 &  7190 & 52 & \ldots \\
ABELL\thin 2029 & 76$^\ddag$ & 350 &  4977 & 77 & (5) \\\hline
\end{tabular}
\end{table}
\begin{figure*}
\centering
\includegraphics[width=4in,angle=270]{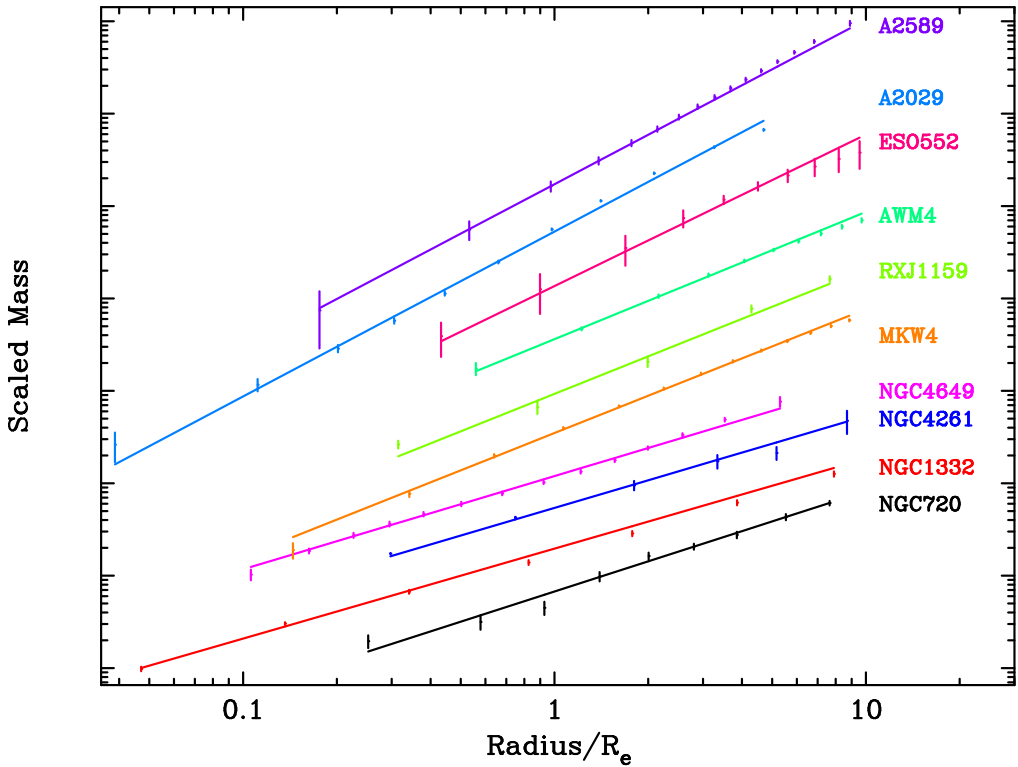}
\caption{Radial mass profiles for each object, arbitrarily scaled for clarity.
The solid lines are the best-fitting profiles determined from our ``forward fitting''
analysis of the temperature and density profiles, while the data-points are 
determined from the more ``traditional'' method (\S~\ref{sect_mass_slope}). We stress that the models are 
{\em not} fitted to these data-points but are derived independently. \label{fig_mass_profiles}}
\end{figure*}
\subsection{The sample}
We chose a sample of 9 objects, spanning $\sim$2 orders of 
magnitude in virial mass (\mvir) from the survey of morphologically relaxed, X-ray luminous systems
by \citet{buote07a}. We supplemented this sample with  
NGC\thin 1332, which has the smallest \reff\ of the relaxed systems we have 
previously studied for X-ray mass analysis \citepalias{humphrey06a,humphrey09d}. 
We focused on \chandra\ data,
since high spatial resolution is essential to study in detail the 
inner part of the X-ray halo. As we require coverage from \ltsim \reff\ to several
times \reff, we only considered objects which had deep enough data to enable spectra to be 
obtained in at least two annuli within \reff. Unfortunately, most nearby 
galaxy clusters which are morphologically relaxed on large scales also exhibit significant 
AGN-induced cavities in the core which complicate the analysis. Therefore, we only included two
clusters which have been shown to be relaxed at both large and small scales 
\citep{lewis02a,zappacosta06a}.
The properties of the sample and the archival
data we used are summarized in Table~\ref{table_sample}.

\subsection{The Slope of the Mass Profile} \label{sect_mass_slope}
 The \chandra\ data were reduced and analysed to provide deprojected temperature and density 
profiles, as outlined in \citetalias{humphrey09d}.
For NGC\thin 1332, NGC\thin 4261 and NGC\thin 4649 the data are described in 
\citetalias{humphrey09d} and \citet{humphrey08a}. 
We will discuss in more detail the NGC\thin 720 data in \citet{humphrey09c} and the A\thin 2029
data in \citet{liu09a}. For RXJ1159+5531, we used the 
deprojected profiles described in \citet[provided by F. Gastaldello 2009, priv comm.]{gastaldello07a}.

Under the hydrostatic approximation, we transformed these density and temperature data into
mass constraints by two complementary approaches.
First, the ``traditional'' method involves parametrizing these profiles
with arbitrary models \citep[for more details on these models, see \citetalias{humphrey09d};][]{humphrey08a,gastaldello07a}, 
which are then differentiated and inserted into the equation of 
hydrostatic equilibrium \citep[\eg][]{mathews78a}. An advantage of this method is that it makes no
{\em a priori} assumption about the form of the mass distribution. 
By evaluating the resulting mass model at a number of radii
(corresponding to each spectral extraction region) we obtained the mass ``data-points'' shown for 
each system in Fig~\ref{fig_mass_profiles}. 
Error-bars were estimated {\em via} a Monte Carlo technique
\citep{lewis03a}. We here focus only on the central part of these data;
based on experimentation, we considered the mass within 10\reff\ or 200~kpc, whichever
is smaller. Over this radial range, the profiles are all approximately powerlaw in form, but the exact
slope varies from object to object.

We find overall good agreement with previously 
published mass profiles \citep[\citetalias{humphrey06a,humphrey09d};][]{gastaldello07a,zappacosta06a,lewis03a},
although for ESO552-020 the normalization is $\sim$0.1~dex higher than that found by \citet{gastaldello07a},
using \xmm\ data. Nevertheless, this discrepancy is comparable with our estimated systematic error for this object 
(\S~\ref{sect_syserr}),
and will not affect our conclusions.

Since the traditional method relies on the adoption of {\em ad hoc} temperature and density profiles,
this can lead to significant systematic errors in the recovered mass distribution \citepalias[\eg][]{humphrey09d}.
Furthermore, the individual ``data-points'' are all correlated, which makes it difficult to 
interpret a fit made directly through them. Therefore, to be more quantitative, we fitted the mass
distribution using the ``forward fitting'' method described in \citet{humphrey08a}.
This involves solving the equation of hydrostatic equilibrium to compute temperature and density profile
models, given parametrized mass and entropy profiles. Since the entropy profile must rise 
monotonically, we parametrized it as a constant plus a powerlaw with one or two breaks added, as needed.
For the stellar plus dark mass distribution, we adopted a powerlaw, corresponding to $\rho_m \propto R^{-\alpha}$,
\ie\
\begin{equation}
M(<R) = M_{75} \left( \frac{R}{\rm 75 kpc}\right)^{3-\alpha}
\end{equation}
where $M$ is the mass enclosed within radius R. $M_{75}$ and $\alpha$ were parameters of the fit.
An additional gas mass component was included self-consistently in the calculation, but is 
generally small in the fitted radial range.
We fitted only the inner parts of the density and temperature profiles, as described above, 
freely varying $\log M_{75}$, $\alpha$, the parameters describing the entropy profile and a term
related to the gas pressure at a suitable reference radius \citepalias{humphrey09d}. Following \citetalias{humphrey09d},
parameter space was explored with a Bayesian
Monte Carlo method\footnote{Specifically, the nested sampling algorithm of \citet{feroz08a}}, assuming flat priors
 (see \S~\ref{sect_syserr} for the impact of using
other priors), and the marginalized best-fitting mass profile parameters and 1-$\sigma$ confidence regions are given 
in Table~\ref{table_results}.  Also shown is the $\chi^2$ per degree of freedom
for the fits to the temperature and density profiles. 
In general, the fits are good. For the three objects for which the fits are not formally
acceptable
(null hypothesis probability $<$5\%), the mean absolute fractional discrepancy between
the temperature and density models and the data are less than $\sim$10\%,
indicating that the true mass distribution is, nevertheless, very close to a powerlaw in form.
For one of these objects (NGC\thin 4261), we did find larger discrepancies within $\sim$0.5~kpc, but
we ignored the data within this region during our fit so as not to bias our results. 
(It is interesting
to note that the full temperature and density profiles of this object 
can be fitted well by a  mass decomposition into luminous and dark matter 
components: \citetalias{humphrey09d}.)
For the two clusters, we have previously published powerlaw fits to the central mass
distributions, based on other data \citep{lewis03a,buote04b,zappacosta06a}, and these studies agree
well with our results.

\begin{figure*}
\centering
\includegraphics[width=6.5in]{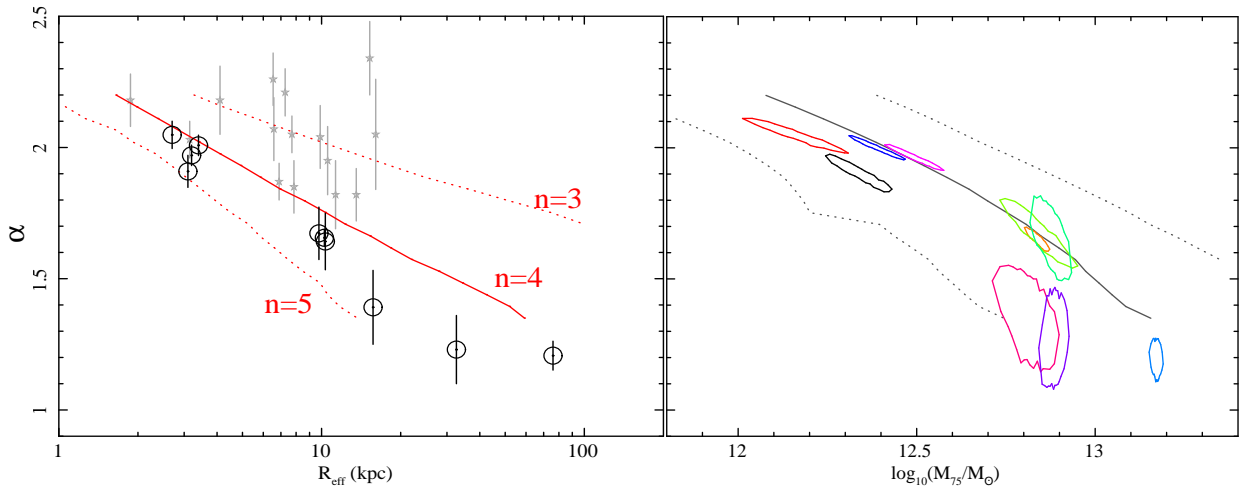}
\caption{{\em Left}: Variation of the marginalized $\alpha$ as a function of \reff\ from
our data (circles),
showing a strong anti-correlation. We overlay (stars) the lensing data of \citet{koopmans06a}, 
which is computed at scales smaller than $\sim$\reff\ and plotted {\em versus} \reff.
For a strictly fair comparison with our work, their data may need to be shifted slightly to the left
(given differences in the adopted photometric bands).
We also overlay the predictions of our toy model (solid line; 
\S~\ref{sect_toy_model}) and an estimate of the model uncertainty (dotted lines). To indicate how changing the Sersic index (n) of the stellar light component
affects our model, we have annotated each model line with the corresponding
value of n.
{\em Right:} Marginalized 1-$\sigma$ joint confidence contours for $\alpha$ and $M_{75}$ for each object.
We overlay the predictions of our toy model, combined with the K-band Kormendy relation (solid line), 
and an estimate of the uncertainty in the model (dotted lines).
\label{fig_mass_slope}}
\end{figure*}
In Fig~\ref{fig_mass_profiles}, we overlay the best-fitting models onto the data-points
obtained from the traditional method, finding overall good agreement. The mean absolute fractional difference
between the model and ``data'' varies from $\sim$5\%\ to $\sim$18\%, and is smaller than 11\%\ for 6 objects.
While we don't expect a pure powerlaw to be an exact description of the mass profile, these residuals
indicate that it is, nevertheless, an excellent approximation to better than $\sim$10--20\%. In fact,
the discrepancies we found between the mass profiles determined from both approaches are comparable
to the typical systematic uncertainties associated with the traditional analysis method 
\citepalias[\eg][]{humphrey09d}, so that the actual agreement between the mass distribution and a powerlaw
shape may be even better.
From Table~\ref{table_results}, it is immediately clear that $\alpha$ systematically varies 
with mass scale. This is shown explicitly in Fig~\ref{fig_mass_slope}, where we plot $\alpha$ {\em versus}
\reff\ and {\em versus} ${M_{75}}$,  in both cases revealing striking anti-correlations.


\begin{table}
\caption{Marginalized best-fitting values of the powerlaw slope ($\alpha$) and normalization
($M_{75}$) for each object in the sample. We list both the marginalized value and the 1-$\sigma$ statistical errors
(``Best'') and an estimate of the systematic uncertainty (``Sys.'') due to various data-analysis choices
(\S~\ref{sect_syserr}). We stress this should {\em not} be added in quadrature with the statistical errors.
Also shown is the $\chi^2$ per degree of freedom (dof) for each fit to the temperature and density profiles.\label{table_results}}
\begin{tabular}{llllll}
Object & \multicolumn{2}{l}{$\alpha$} & \multicolumn{2}{l}{$log M_{75}$} & $\chi^2$/dof \\
 & Best & Sys. & Best & Sys. & \\ \hline
NGC\thin 1332& $2.05^{+0.05}_{-0.06}$&$\pm 0.04$& $12.15^{+0.12}_{-0.11}$&$^{+0.17}_{-0.07}$ & 16/6 \\
NGC\thin 720& $1.91^{+0.05}_{-0.07}$&$\pm 0.02$& $12.34^{+0.08}_{-0.07}$&$^{+0.03}_{-0.03}$ & 5.7/8 \\
NGC\thin 4649& $1.97^{+0.03}_{-0.05}$&$\pm 0.03$& $12.48^{+0.08}_{-0.05}$&$^{+0.23}_{-0.09}$ & 30/22 \\
NGC\thin 4261& $2.01^{+0.03}_{-0.05}$&$^{+0.09}_{-0.04}$& $12.37^{+0.09}_{-0.04}$&$\pm 0.02$ & 6.5/2 \\
RXJ1159& $1.67^{+0.11}_{-0.10}$&$\pm 0.06$& $12.82^{+0.10}_{-0.05}$&$+0.07$ & 9.3/4 \\
MKW4& $1.66^{+0.03}_{-0.04}$&$\pm0.03$& $12.84^{+0.03}_{-0.02}$&$^{+0.06}_{-0.02}$ & 26/18 \\
AWM4& $1.64^{+0.13}_{-0.10}$&$\pm 0.08$& $12.88^{+0.03}_{-0.05}$&$^{+0.11}_{-0.06}$ & 15/12 \\
ESO552& $1.39^{+0.11}_{-0.18}$&$^{+0.09}_{-0.07}$& $12.81^{+0.07}_{-0.06}$&$+0.10$ & 9.6/10 \\
A\thin 2589& $1.23^{+0.17}_{-0.10}$&$^{+0.27}_{-0.10}$& $12.88^{+0.03}_{-0.03}$&$\pm 0.12$ & 19/17 \\
A\thin 2029& $1.21^{+0.05}_{-0.06}$&$^{+0.22}_{-0.11}$& $13.17^{+0.01}_{-0.01}$&$\pm 0.11$ & 23/8 \\ \hline
\end{tabular}
\end{table}

\subsection{Systematic Errors} \label{sect_syserr}
As in all studies of dark matter in early-type galaxies, regardless of the specific method, our work
involved a number of arbitrary analysis choices. In this section, we describe how sensitive our
results are to the various choices we made. 
For a more detailed discussion of these various 
systematic error assessments, see \citetalias{humphrey09d}. 

We first examined how the choice of priors might be influencing our results by replacing each of the
flat priors we adopted by priors which were flat in logarithmic space (we replaced the 
flat prior on $\log M_{75}$ with one which is flat on $M_{75}$).
These choices typically had a smaller effect than the statistical errors. Next we investigated the sensitivity of our spectral-fitting results to our treatment of the \chandra\ background 
by using the standard ``background template'' spectra, suitably renormalized to match the 
data at energies \gtsim 10~keV, instead of the more robust, modelled background
adopted by default. To assess the importance of the radial range  we fitted, we tried reducing it by
$\sim$20\%\ for each object. 
We experimented with spectral-fitting over different energy ranges (0.4--7.0~keV,
0.5--2.0~keV and 0.7--7.0~keV, in addition to our default, 0.5--7.0~keV), varying the neutral galactic column
density by 25\%, and the distance by 30\%\ (for the NGC objects, we instead varied the distance by the 
statistical error on their distance measurements in \citealt{tonry01}).
In Table~\ref{table_results} we list the largest change in the marginalized parameter values arising due to 
these choices. None of these systematic errors are large enough to affect our conclusions.


\section{Discussion}
\subsection{A luminous-dark matter conspiracy}
\begin{figure}
\includegraphics[width=2.3in,angle=270]{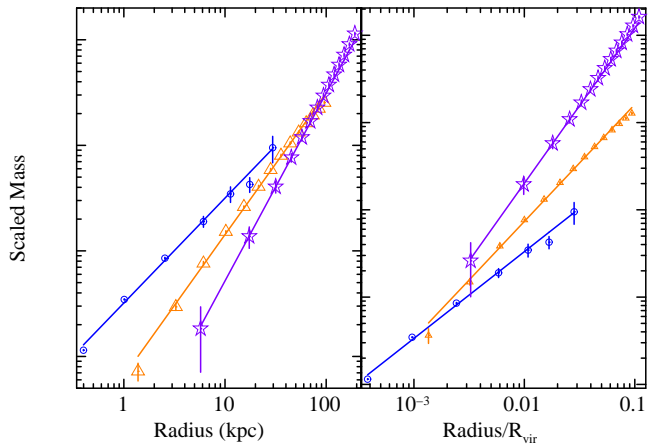}
\caption{{\em Left:} Comparison of the mass profiles of NGC\thin 4261 (circles), MKW4 (triangles) and A\thin 2589 (stars),
normalized to have the same mass at 100~kpc. Note that the shapes of the mass profiles are different even over overlapping radial ranges.
{\em Right:} The same, but scaling the radius with respect to the virial radius
($R_{vir}$) and renormalizing each profile to have the same mass at 
0.001$R_{vir}$.\label{fig_mass_profiles2}}
\end{figure}
In general, the stars are believed to dominate the gravitational potential within $\sim$\reff,
with the dark matter dominating outside this scale \citep[\eg][]{brighenti97a,gerhard01a};
in particular, we have shown this for the systems studied here 
\citep[\citetalias{humphrey06a,humphrey09d}][]{gastaldello07a,lewis03a,buote04b,zappacosta06a}.
Therefore,  the approximately powerlaw mass distributions
shown in Fig~\ref{fig_mass_profiles} indicate an apparent conspiracy between the luminous
and dark matter to produce a scale-free total mass distribution, at least within 
$\sim$10\reff. At much larger scales, there is evidence that the profiles deviate from this 
simple shape \citep[\eg][]{lewis03a,vikhlinin06b,humphrey06a,gastaldello07a,romanowsky09a}. 
Nevertheless, in the inner
regions of the systems, this effect is
similar to the ``bulge-halo'' conspiracy found at much smaller
scales (\ltsim\reff) in lensing and stellar dynamics studies 
\citep[\eg][]{treu04a,koopmans06a,koopmans09a}.
{It is important to appreciate that the observed powerlaw mass profiles cannot
simply arise from the arbitrary superposition of any reasonable stellar and dark
mass components. To illustrate this quantitatively, we simulated a powerlaw-like 
mass profile and fitted it with a model comprising
 Sersic and dark matter components (as described in detail in \S~\ref{sect_toy_model}).
For any 
given galaxy luminosity, we found that both \reff\ and the normalization of the dark matter component
cannot vary by more than $\sim$25\%\ from their ``best'' values while maintaining mean absolute
fractional deviations between the model and data of less than 10\%\ (as indicated by our fits in 
\S~\ref{sect_mass_slope}).}

The origin of this conspiracy is unclear, but it is likely tied to the complex
interaction of baryons and dark matter in the centres of galaxies
(for example, \citealt{robertson06b} point out the importance of gas physics in maintaining
the tilt of the fundamental plane, which may be related to this conspiracy;
\S~\ref{sect_fp}). Unfortunately these processes are very poorly understood;
the predicted central dark matter density cusps due to ``adiabatic
contraction'' \citep{blumenthal86a,gnedin04a} have not been observationally
confirmed \citep[\eg][]{humphrey06a,zappacosta06a,gnedin07a,dutton07a}, suggesting that
other effects, such as dynamical friction, may be important 
\citep[\eg][]{elzant04a,dutton07a,abadi09a}. It is unclear, therefore, that
any model is so far robustly predicting a powerlaw {\em total} mass distribution
arising from these effects.

Since the data were not all fitted over the same physical scales, it is important
to assess whether the systematic variation in $\alpha$ we observe
is simply an artefact of these different fit ranges. In Fig~\ref{fig_mass_profiles2}
we show the mass profiles of three representative systems with different \reff,
arbitrarily scaled for clarity, and shown as a function of physical radius
(left)
and fraction of the virial radius \citep[right; deduced from the virial masses
reported in][]{buote07a}. In conjunction with Fig~\ref{fig_mass_profiles}
(which shows the same profiles as a function of fraction of \reff), it is immediately
clear that {\em throughout overlapping radial scales} the slopes of the three mass
distributions are all very different, so that the fit range is unlikely to be the cause
of the trends observed in Fig~\ref{fig_mass_slope}.

Although all of the systems we considered are morphologically relaxed,
which should eliminate objects with the largest departures from
hydrostatic equilibrium, it is still important to consider whether
deviations from hydrostatic equilibrium could be responsible for the
observed trends.
In order to affect the {\em shape} of
the mass profile, any source of nonthermal pressure must vary radially
in a finely balanced manner, given that the gas pressure profile
is consistent with a powerlaw mass density profile (Fig~\ref{fig_mass_profiles})
that varies systematically with the host properties (Fig~\ref{fig_mass_slope}).
For the cluster A\thin 2589,
\citet{zappacosta06a} concluded that, 
if the true mass distribution is much cuspier than inferred from the 
X-rays (for example, an unmodified NFW plus a stellar mass component),
 no plausible source of turbulent or magnetic pressure could make the 
X-ray measurements imply such a flat total density profile.
\citetalias{humphrey09d} carried out full mass decompositions for
three of our galaxies, finding stellar mass to light ratios 
in excellent agreement with the predictions of stellar population 
synthesis models, suggesting that nonthermal pressure is very small
(\ltsim 10--20\%) within \reff.
This was further supported by the good agreement between the X-ray 
constraints on the central black hole masses, and those obtained 
from optical methods for these objects. Furthermore, 
nine of our objects were included in the 
sample of \citet{buote07a}, who showed that their dark
matter halos inferred from X-ray mass analysis approximately exhibit the 
relationship between halo ``concentration'' (\cvir) and total mass predicted
from numerical structure formation simulations. This supports the 
idea that the shapes of the mass profiles for these 
systems are not significantly  in error due to non-hydrostatic
effects.

In their survey of 15 lensing
systems, \citet{koopmans06a} found that $\alpha$ ranged from $\sim$1.8--2.4,
while the \reff\ spanned $\sim$2--17~kpc. In the left hand panel
of Fig~\ref{fig_mass_slope}, we overlay their measured $\alpha$ {\em versus}
\reff. We find that they
are roughly consistent with the points from our X-ray sample, but at
given \reff, $\alpha$ is generally larger. For strict consistency with our 
adopted photometric bands, their data may need to be adjusted, however.
Since \reff\ in the K-band is typically smaller than in bluer bands 
(as an illustration of this 
effect, for the galaxies in the catalogue of \citealt{labarbera08a}, we 
find that that \reff\ is on average $\sim$50\%\ larger in the R-band than
the K-band), this would likely involve shifting the \citeauthor{koopmans06a}
to the left, bringing them into better 
agreement with our results. By inspection, there is  a slight hint
in their data  of the anti-correlation we observe between $\alpha$ and
\reff, although it is not statistically significant. 
However, since their fits were all restricted to
regions \ltsim\ 0.9\reff, where the stellar mass dominates the potential,
they are more likely to have been subject to local curvature in the mass distribution than 
the X-ray data were, given the better radial coverage of the latter.

\subsection{Decomposing the mass profiles: a toy model} \label{sect_toy_model}
\begin{figure}
\includegraphics[width=2.5in,angle=270]{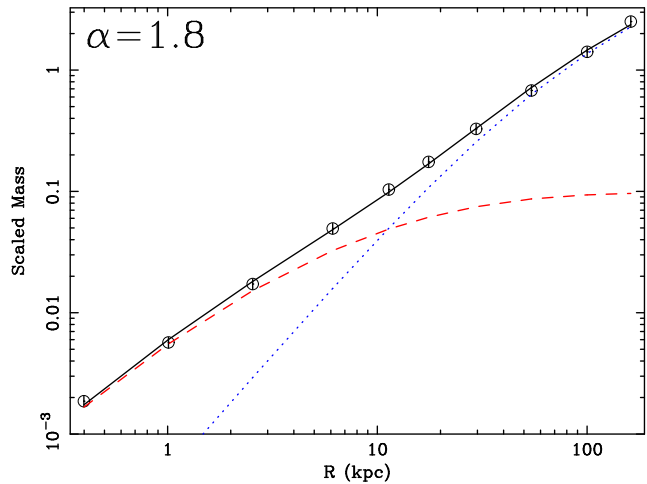}
\caption{Sample simulated mass profile and associated
mass decomposition. The data-points correspond to a powerlaw mass distribution
with $\alpha=1.8$, while the solid line indicates the best-fitting two-component (dark plus luminous 
matter) model. Error-bars, corresponding to a 10\%\ fractional uncertainty, are shown to guide the eye.
Also shown are the corresponding best-fitting dark matter (dotted line) and stellar mass
(dashed line) models. Note how the sum of two scale-dependent models produces an approximately powerlaw
total mass distribution over the fitted range.\label{fig_mass_decomposition}}
\end{figure}
To investigate quantitatively the implications of the apparent conspiracy between dark and luminous 
mass components implied by Fig~\ref{fig_mass_profiles}, we constructed a simple
toy model by decomposing a pure powerlaw mass distribution into an NFW \citep{navarro97} dark matter
profile plus a de Vaucouleurs stellar mass component
\citep[using the deprojected Sersic approximation of][and fixing the Sersic index, $n=4$]{prugniel97a}. 
{In our previous work (\citealt{buote07a}; \citetalias{humphrey06a,humphrey09d}; 
\citealt{gastaldello07a,zappacosta06a}), we have shown that decompositions of this type for the systems
studied in the present work
give parameters for the NFW model which are consistent with theory. The measured NFW concentration
parameters are slightly higher than predicted median from N-body simulations, but this can be reasonably
attributed to relaxed X-ray selected systems having formed early}.
To perform the decomposition,
we first constructed a set of mass ``data-points'' from the powerlaw model 
in 10 approximately logarithmically-spaced bins
between $\sim$0.4--200~kpc, which we then fitted with the two-component model
by minimizing the rms of the fractional residuals.
The normalization and scale radius of the dark matter component 
were allowed to vary freely, as were \reff\ and the normalization of the stellar model. 
We allowed the slope of the input mass distribution to vary systematically from $\alpha=2.2$ to 
$\alpha=1.35$. Below this limit, we found that the derived relationships (discussed below)
became non-monotonic, which may represent the limitations of our simple parametrization 
rather than any physical insight, and so we prefer not to discuss this regime.
While one might {\em a priori} expect significant degeneracies in
this decomposition given the number of free parameters, in fact the shapes
of the Sersic and NFW mass profiles are sufficiently different that we 
found a unique decomposition for any input powerlaw, 
at least for the range of profiles and models we investigated.
We show an example
simulated profile, and the best-fitting dark and stellar mass components, in Fig~\ref{fig_mass_decomposition},
clearly illustrating how they can conspire to produce an approximately scale-free model. 

We overlay in Fig~\ref{fig_mass_slope} (left panel) 
the best-fitting relation between \reff\ and $\alpha$ obtained
from our simple mass decompositions (solid line), which can be approximated to better than
1\%\ by $\alpha = 2.31-0.54\log R_e$.
Without any fine tuning, it clearly captures the overall behaviour
of the observed data reasonably well. Since the stellar light profiles of real galaxies may not be 
exactly de Vaucouleurs in form, we have also experimented with varying the Sersic index, n, from 3--5.
Similarly, the NFW model used to parameterize the dark matter halo over the fitting range and may not
be strictly correct if the baryonic and non-baryonic components interact gravitationally.
Therefore, we have tried replacing it with the revised mass model of  \citet{navarro04a},
allowing the mass slope index to fit freely and thus 
provide more freedom in parameterizing the radial distribution of dark matter. 
The dotted lines in Fig~\ref{fig_mass_slope}
bound the range of recovered relations, given these changes, which clearly bracket the observed data. In practice, the impact of changing the stellar 
light profile dominated this estimate of our model uncertainty.

Since our simple toy model can be freely renormalized, we cannot use it alone to make
any explicit predictions for the relationship between $\alpha$ and $M_{75}$. One way to make
further progress, however, is to exploit the ``Kormendy relation'' between a galaxy's luminosity
and \reff\ \citep{kormendy77a}. For a given stellar mass-to-light ratio ($\gamma_*$),
$M_{75}$ must be adjusted  to keep the stellar component consistent with the Kormendy relation
while maintaining the powerlaw total mass distribution for a given $\alpha$ (and hence \reff).  
We adopted a form for this relation which is appropriate for
our data, estimated from the K-band galaxy luminosities and \reff\  
tabulated by \citet{gastaldello07a} and \citetalias{humphrey06a}.
We found a mean relation $\log L/L_\odot = 10.95 + 0.79 \log R_e$, with
an intrinsic scatter of $\sim$0.11~dex, 
where $L$ is the total luminosity of the galaxy
in the K-band and $L_\odot$ is the luminosity of the Sun, assuming a K-band absolute Solar magnitude of 
3.41. We assumed $\gamma_*=1$, appropriate for an old stellar population 
\citep[\eg][]{maraston05a}.
In Fig~\ref{fig_mass_slope} (right panel)
we overlay the predictions of this simple model (solid line), and an estimate of the range of 
uncertainty, which factors in the effects of varying the Sersic index of the stellar
light by $\pm 1$, the intrinsic scatter in the Kormendy relation and a $\pm$50\%\ variation
in $\gamma_*$ (which dominates the error budget). Clearly the model
captures the overall behaviour of the data fairly well.

{To verify that our model is self-consistent, it is important to determine if the parameters of the 
fitted dark matter model it predicts are reasonable. Since the toy model 
cannot be used alone to set the normalization of the mass profile, to do this we considered the 
simple model, folding in the Kormendy relation, described above. Starting with the 
$\alpha$-$M_{75}$ relation predicted by this model, we simulated corresponding powerlaw 
mass profiles, and decomposed them into Sersic and dark matter components. We found monotonic
relations between\ \mvir\ and $\alpha$ and between NFW \cvir\ and $\alpha$, allowing
us to recast this as a \cvir-\mvir\ relation, which we can approximate by 
\cvir$=c_{14} \left( M_{vir}/10^{14} M_\odot\right)^{-\beta}$, where $c_{14} = 10$ and 
$\beta=0.17$. This is very close to the best-fitting \cvir-\mvir\ relation found
by \citet{buote07a}, for which $c_{14}=9.0\pm0.4$ and $\beta=0.172\pm0.026$. Exploring the 
estimated  range of uncertainty on the model linking $\alpha$ and $M_{75}$, we found
$c_{14}$ to lie in the range 6--16 and $\beta$ from 0.13--0.18.  The total range of \mvir\
spanned by our predicted \cvir-\mvir\ relation ($\sim 12.5<log_{10} M_{vir} < 15.0$) is similar
to the range spanned by the systems in our sample. It appears, therefore, that the parameters of the 
NFW component implied by our simple model are appropriate for the kinds of system studied in this work.
}

\subsection{Implications for the fundamental plane} \label{sect_fp}
\begin{figure}
\includegraphics[width=2.5in,angle=270]{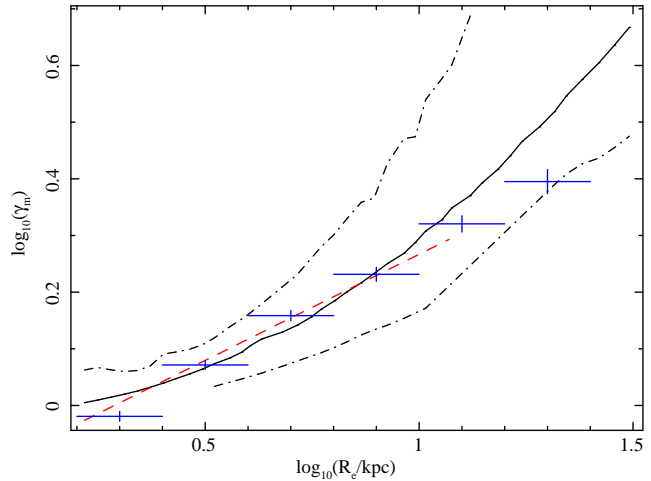}
\caption{The predicted ratio between the total mass and 
stellar mass enclosed within \reff\ from
our toy model  ($\gamma_m$), shown as a function of \reff\ (solid line). 
The dot-dash lines indicate our estimate of the uncertainty on the 
model, and the dashed line indicates the best-fitting linear approximation in the range
$\log R_e=$0.2--1.0. The data-points are the observed K-band M/L ratios from the 
data of \citet{labarbera08a} (see text).
Note the good agreement between the data and the toy model.
\label{fig_mass_to_light}}
\end{figure}
The existence of a conspiracy between the dark and luminous matter to 
produce a powerlaw  {\em total} mass profile should have implications for the 
tilt of the fundamental plane. The fundamental plane may be written in the form
\begin{equation}
\log R_e = A \log \sigma_0 + B \mu + const.  \label{eqn_fp}
\end{equation}
where  $\sigma_0$ is the central velocity dispersion,
$\mu$ is the mean surface brightness within \reff, and A and B
are proportionality constants. One way to gain insight into this relation is
to interpret it in terms of the scalar virial theorem \citep[\eg][]{binney08a}.
We can write this in the form:
\begin{eqnarray}
\log R_e & = & 2\log \sigma_0 - \log\left(\frac{L_{1/2}}{\pi R_e^2}\right) - \log \gamma + 5.47 \nonumber
\end{eqnarray}
Here \reff\ is in kpc, $\sigma_0$ is in $km\ s^{-1}$,
$L_{1/2}$ is (approximately) 
the light within \reff\ (in Solar units) and  $\gamma = \gamma_* \gamma_m$  is the mass to light
(M/L) ratio in Solar units, where $\gamma_*$ is the stellar M/L ratio and $\gamma_m$ the 
ratio of total mass 
to stellar mass within \reff. {The value of the additive constant on the right hand side is useful only
in setting the absolute normalization of $\gamma_*$ and does not affect our conclusions.
For convenience we chose to determine it from the theoretical 
relation of \citet{wolf09a}\footnote{We replaced the total velocity 
dispersion in their relation with the central velocity dispersion, which is only strictly
valid for a flat velocity dispersion profile.}, 
since this implies a mean $\gamma_*$ of $\sim$1 for the \citet{labarbera08a} galaxy 
sample (see below), which is consistent with the expectations for an old stellar population.
Alternative, empirical relations have been found by 
\citet[][assuming mass follows light]{cappellari06a} and  
\citet[][assuming a singular isothermal sphere mass distribution]{bolton07a},
which would imply lower $\gamma_*$, by as much as $\sim$40\%\ (for the \citeauthor{cappellari06a}
relation). Nevertheless, for our work the absolute value of $\gamma_*$ is unimportant, and so
this choice does not affect our conclusions.}


In the K-band, assuming the absolute Solar magnitude
is 3.41 and neglecting cosmological effects, we can write 
$\log (L_{1/2}/\pi R_e^2) \simeq 16.00 -0.4\mu$, and hence
\begin{equation}
\log R_e  = 2\log \sigma_0 +0.4 \mu -\log \gamma - 10.53 \label{eqn_virial}
\end{equation}
If structural non-homology does not significantly affect this equation, 
the tilt of the fundamental plane
 arises through the dependence of $\gamma$ on the other properties of the
galaxy. Eliminating $\sigma_0$ between Eqns~\ref{eqn_fp} and \ref{eqn_virial}
 and rearranging, we obtain:
\begin{eqnarray}
\log \gamma & = & \left( \frac{2}{A}-1 \right) \log R_e - \left( \frac{2B}{A} - 0.4 \right) \mu + const. \nonumber \\
 & = & (0.33\pm0.04) \log R_e + (0.007\pm0.011) \mu + const. \nonumber\\
 & \simeq &  (0.33\pm0.04) \log R_e + const. \label{eqn_gamma}
\end{eqnarray}
where we have used the ``$\log \sigma_0$ fit'' values for A and B from the recent 
K-band analysis of \citet{labarbera08a}. The $\mu$ term is consistent with zero,
within the error, suggesting that the tilt of the fundamental plane can arise if 
$\gamma$ depends only on \reff, which  is exactly the prediction of the 
toy model, as we discuss below. Nevertheless, since the $\mu$ term may be non-negligible,
to obtain a more accurate assessment of how $\gamma$ varies with \reff, we 
computed $\log \gamma$ directly from the data of \citeauthor{labarbera08a}
by using Eqn~\ref{eqn_virial}. We grouped these data into a series of narrow \reff\ bins\footnote{for consistency with our X-ray analysis, we ignored those systems with $\log$\reff \ltsim 0.2},
within each of which we calculated the mean $\gamma$ and its error (by applying the 
central limit theorem to the scatter of the data), as shown in Fig~\ref{fig_mass_to_light}.

We can use our toy model not only to predict the relationship between \reff\ and $\alpha$,
but also the dark matter fraction within \reff\ (and hence $\gamma_m$), as is immediately clear from inspection of
Fig~\ref{fig_mass_decomposition}. Since the dark matter {\em fraction} does not depend on the 
overall normalization of the mass profile, we do {\em not} need to fold in any other 
relations (such as the Kormendy relation). Furthermore, since we find that \reff\ varies
monotonically with $\alpha$ (Fig~\ref{fig_mass_slope}), 
it is possible to recast this as a relation between $\gamma_m$ and \reff. We show the resulting
relation (roughly $\log \gamma_m \simeq -0.03\log R_e +0.32 \left( \log R_e \right)^2$) in Fig~\ref{fig_mass_to_light}, along with an estimate of the 
uncertainty that arises from from varying the Sersic index by $\pm$1, or adopting the 
\citet{navarro04a}
model for the dark matter profile. It is immediately clear that there is good
agreement between the model and the \citet{labarbera08a} data, {implying the mean $\gamma_* \simeq 1$,
as noted above}. Since the 
stellar populations of early-type galaxies (and hence $\gamma_*$) do not strongly 
vary with \reff\ \citep[at least for \reff \gtsim 1~kpc:][]{graves09b}, this
{suggests that assuming the ubiquity of approximately powerlaw density profiles (supported 
by Fig~\ref{fig_mass_profiles}) is sufficient
(albeit not necessary) to explain the tilt of the fundamental plane. While it is possible that 
not all giant early-type galaxies have density profiles which so closely 
resemble a powerlaw (although we are not aware of any convincing such counter-examples found by X-ray methods), 
to lie on the fundamental plane, Fig~\ref{fig_mass_to_light} indicates that their $\gamma_m$ must still
vary with \reff\ in a very similar way to the predictions of the toy model.
Nevertheless, since it is actually the flattening of the density profile
with increasing \reff\ that is responsible for the $\gamma_m$-\reff\ relation
predicted by the toy model, it is plausible that similar flattening of 
the mass distributions in a system without a powerlaw-like mass distribution might produce approximately
the same trend.}

One intriguing prediction of our toy model is that the relationship between $\log \gamma_m$ 
and $\log R_e$ has some curvature, which would also imply curvature in the fundamental
plane. There is, in fact, some evidence that brightest cluster galaxies (BCGs) may not 
lie on the same fundamental plane as smaller systems \citep[\eg][]{oegerle91a,vonderlinden07a,bernardi07a}, but it is not clear that the observed behaviour could be explained in 
terms of the predicted curvature. Indeed, there is a hint that the observed $\gamma$ {\em versus}
\reff\ relation begins to deviate from our simple predictions at \reff\gtsim 20~kpc
(Fig~\ref{fig_mass_to_light}), which may arise due to deviations from structural homology among
BCGs, or it may represent the limitations of our simple model.

To make a more quantitative comparison between the traditional fundamental plane and our toy 
model, we  adopted a linear approximation for the toy model relationship
between $\gamma_m$ and \reff\ ($\log \gamma_m \simeq -0.11 +0.37 \log R_e$;
this is accurate to about $\sim$0.03~dex over the range $0.2<\log R_e<1.0$, as shown
in Fig~\ref{fig_mass_to_light}, assuming $\gamma_*=1$). Substituting this into 
Eqn~\ref{eqn_virial} (assuming $\gamma_*$ is constant) and 
rearranging into the form of  of Eqn~\ref{eqn_fp}, we found  A=1.46 and B=0.29. 
These are in good agreement with recent K-band measurements of the 
fundamental plane \citep[A$\sim$1.4--1.5 and B$\sim$0.30--0.33:][]{labarbera08a,jun08a,mobasher99a,pahre95a}.
Repeating the analysis, but allowing the Sersic index to vary by $\pm$1,
or using the alternative dark matter model of \citet{navarro04a}, 
we obtained A in the range $\sim$1.2--1.6 and 
B$\simeq$0.23--0.31.  Thus, assuming only a powerlaw total mass distribution, we can
clearly reproduce the shape of the K-band fundamental plane, without any fine-tuning
of the model parameters.  We therefore speculate that 
the establishment of a powerlaw total mass distribution is a fundamental feature of galaxy 
formation and the principal factor in determining the tilt of the fundamental plane.
Since the relative importance of gas-rich and ``dry'' merging for the assembly of present-day
giant ellipticals is still under debate \citep[\eg][]{hopkins08a,naab09a,nipoti09a}, 
the requirement of producing a nearly powerlaw total mass distribution in the core of low-redshift 
systems will therefore be a key test of these models.

\section*{Acknowledgments}
We thank Fabio Gastaldello and Wenhao Liu for making available their deprojected density and 
temperature profiles. We thank Bill Mathews, Chris Fassnacht and Luca Ciotti for stimulating 
and helpful discussions.
We also thank Fabio Gastaldello {and Michele Cappellari} for providing comments on a draft of the paper.
This research made use of the
NASA/IPAC Extragalactic Database (\ned)
which is operated by the Jet Propulsion Laboratory, California Institute of
Technology, under contract with NASA.
Partial support for this work was provided by NASA under 
grant NNG04GE76G issued through the Office of Space Sciences Long-Term
Space Astrophysics Program. 

\bibliographystyle{mnras_hyper}
\bibliography{paper_bibliography2.bib}

\end{document}